\documentclass[conference]{IEEEtran}
\usepackage{cite}

\ifCLASSINFOpdf
\usepackage[pdftex]{graphicx}
\else
\fi
\usepackage{amsmath}
\usepackage{amsfonts} 
\usepackage{amsthm}
\usepackage{amssymb}
\usepackage[boxruled]{algorithm2e}
\usepackage{amsthm} 
\usepackage{mathtools} 

\usepackage[T1]{fontenc}  
\usepackage{textcomp}     
\usepackage{mathptmx}     

\usepackage{BOONDOX-ds}

\usepackage{bbm}

\newcommand{\pfun}{\mathop{\hbox{$\to$\kern-7pt\raise.9pt\hbox{\scalebox{1}[.55]{$|$}}\kern4pt} }}

\usepackage{array}

\begin{document}

\title{Optimizing Xeon Phi for Interactive Data Analysis}

\author{\IEEEauthorblockN{Chansup Byun$^1$, Jeremy Kepner$^{1,2,3}$, \\ 
William Arcand$^1$, David  Bestor$^1$, William Bergeron$^1$,  Matthew Hubbell$^1$, Vijay Gadepally$^{1,2}$, \\ 
Michael Houle$^1$, Michael Jones$^1$, Anne Klein$^1$,  Lauren Milechin$^4$, Peter Michaleas$^1$, \\ 
Julie Mullen$^1$, Andrew Prout$^1$, Antonio Rosa$^1$,  Siddharth Samsi$^1$, Charles Yee$^1$, Albert Reuther$^1$
\\
\IEEEauthorblockA{$^1$MIT Lincoln Laboratory Supercomputing Center, $^2$MIT Computer Science \& AI Laboratory, \\ $^3$MIT Mathematics Department, $^4$MIT Department of Earth, Atmospheric and Planetary Sciences}}}
\maketitle

\begin{abstract}
The Intel Xeon Phi manycore processor is designed to provide high performance matrix computations of the type often performed in data analysis.  Common data analysis environments include Matlab, GNU Octave, Julia, Python, and R.  Achieving optimal performance of matrix operations within data analysis environments requires tuning the Xeon Phi OpenMP settings, process pinning, and memory modes.  This paper describes matrix multiplication performance results for Matlab and GNU Octave over a variety of combinations of process counts and OpenMP threads and Xeon Phi memory modes.  These results indicate that using {\sf\small KMP\_AFFINITY=granlarity=fine}, {\sf\small taskset} pinning, and {\sf\small all2all cache} memory mode allows both Matlab and GNU Octave to achieve 66\% of the practical peak performance for process counts ranging from 1 to 64 and OpenMP threads ranging from 1 to 64. These settings have resulted in generally improved performance across a range of applications and has enabled our Xeon Phi system to deliver significant results in a number of real-world applications.
\end{abstract}

%
\IEEEpeerreviewmaketitle

\section{Introduction}
\let\thefootnote\relax\footnotetext{This material is based upon work supported by the Assistant Secretary of Defense for Research and Engineering under Air Force Contract No. FA8702-15-D-0001 and National Science Foundation grants DMS-1312831 and CCF-1533644. Any opinions, findings, conclusions or recommendations expressed in this material are those of the author(s) and do not necessarily reflect the views of the Assistant Secretary of Defense for Research and Engineering or the National Science Foundation.}

\begin{figure}[t]
\centering
\includegraphics[width=\columnwidth]{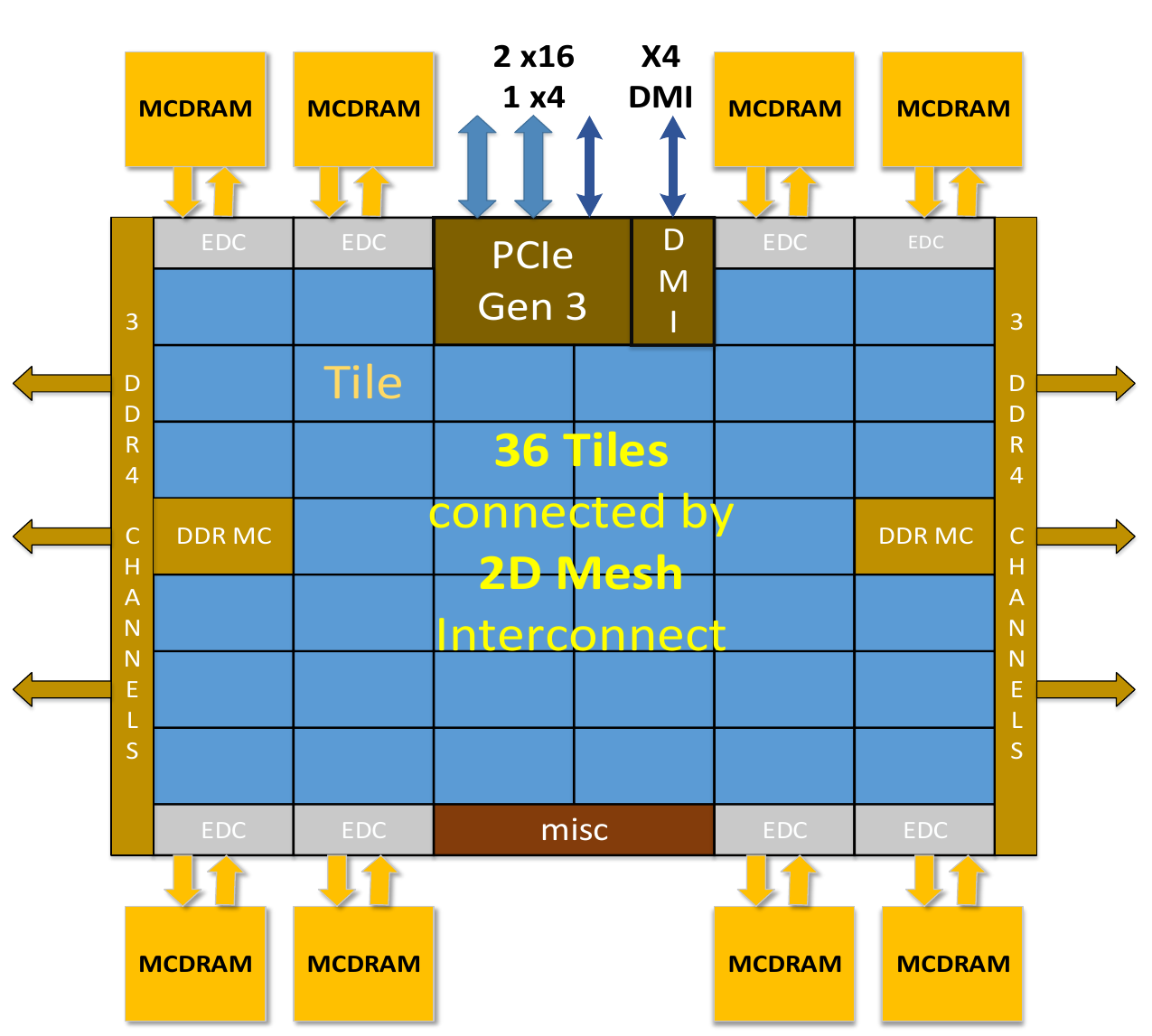}
\caption{Xeon Phi 36 tile layout from \cite{sodani2015knights}. Xeon Phi processors ship with different numbers of tile enabled: 36 tile (Xeon Phi 7250), 34 tile (Xeon Phi 7230), and 32 tile (Xeon Phi 7210 - this paper).}
\label{fig:XeonPhiLayout}
\end{figure}

\begin{figure}[t]
\centering
\includegraphics[width=0.5\columnwidth]{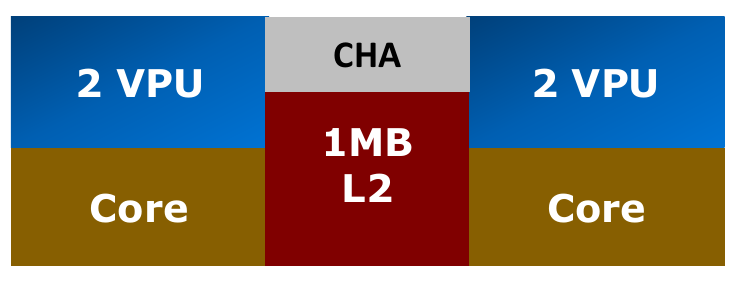}
\caption{Xeon Phi tile structure from \cite{sodani2015knights}.  Each tile has two cores.  Each core has two virtual processors, two AVX512 vector math units, and four hyperthreads.}
\label{fig:XeonPhiTile}
\end{figure}

The Intel Xeon Phi 72x0 (KNL - Knights Landing) processor represents an important contribution in a long-line  of manycore processors \cite{mcmahon1996space,taylor2002raw,mattson2008programming,ramey2011tile} with high-core count ($\geq$64), large number of vector units ($\geq$128), tiled physical layout, and  high speed memory combined with significant amounts of DRAM \cite{sodani2015knights,sodani2016knights} (see Figures~\ref{fig:XeonPhiLayout} and \ref{fig:XeonPhiTile}). The Xeon Phi is ideally suited to applications that perform many vector operations.  Matrix multiplication is a common data analysis operation \cite{kepnerjananthan} that is well-suited to the Xeon Phi processor.  Mathematically matrix-matrix multiplication is denoted
$$
  \mathbf{C} = \mathbf{A} \mathbf{B}
$$
where $\mathbf{A}$ is a N$\times$L matrix, $\mathbf{B}$ is a L$\times$M matrix, and $\mathbf{C}$ is a M$\times$N matrix.  

Increasingly, data analysis is performed in high-level programming environments that include Matlab, GNU Octave, Julia, Python, and R.  These environments allow a programmer to invoke the full power of a processor such as the Xeon Phi with simple, intuitive syntax
$$
  {\sf\small  C = A*B}
$$
While the above code makes matrix multiplication easy to invoke, there are significant additional tuning and configuration steps necessary to allow such an operation to achieve maximum performance \cite{dongarra2015hpc,doerfler2016roofline,haidar2016lu,jeffers2016intel,chunduri2017analytical,nagasaka2018high,lim2018implementation}.  These steps are often outside the domain of expertise of data analysis programmers and best provided by systems operators.  The Lincoln Laboratory Supercomputing Center (LLSC) operates a 648-node Xeon Phi supercomputer.  Our focus is on interactive high performance environments so this work explores the steps necessary to allow these environments (Matlab and GNU Octave specifically) to achieve maximum performance on matrix multiplication as invoked by the above Matlab/Octave code syntax.

Our prior work has focused on the interactive launch of thousands of data analysis environments across hundreds of nodes \cite{kepner2018design,gadepally2018hyperscaling,jones2018interactive,reuther2018interactive, kepner2018tabularosa}.  This paper focuses on the various methods we used to get maximum single node Xeon Phi performance.  In particular with respect to OpenMP parameters, process pinning, and memory settings.  We have found these settings have resulted in generally improved performance across a range of applications and has enabled our Xeon Phi system to deliver significant results that have enabled a number of real-world applications in health sciences \cite{trafton2017mapping}, hurricane relief \cite{foy2018lidar}, astronomy \cite{lindsay2018using}, and cybersecurity \cite{mcgovern2019supercomputers}. The rest of the paper is organized as follow.  First, the effective OpenMP parameters for Matlab and GNU Octave are given. Second, the method for pinning processes to cores is presented.  Third, the Xeon Phi memory modes are described.  Finally, the integrated overall performance measurements are presented for the different memory modes.

\section{OpenMP}

OpenMP (www.openmp.org) is an application programming interface that supports multi-platform shared memory multiprocessing programming in C, C++, and Fortran.  OpenMP is an important tool used in many math libraries to exploit multiple cores on a shared memory compute node.  The maximum parallelism that OpenMP will seek to exploit is often set via the environment variable {\sf\small OMP\_NUM\_THREADS}.

To allow a user to readily control the number of nodes, processes, and OpenMP threads their parallel Matlab/Octave program uses, the LLSC system uses our pMatlab \cite{kepner2009parallel} manycore launch infrastructure and its simple interactive parallel launch syntax
$$
  {\text {\sf\small pRUN({\textquotesingle}MyCode{\textquotesingle},[Nnode Nproc Nthread],{\textquotesingle}system{\textquotesingle})}}
$$
In the above syntax {\sf\small Nnode} is the number of compute nodes that the user desires to run on, {\sf\small Nproc} is the number of processes (distinct Matlab/Octave instances) per node, and {\sf\small Nthread} sets the value of {\sf\small OMP\_NUM\_THREADS}.  In this paper, the focus is on single node performance ({\sf\small Nnode}=1) and the number processes and OpenMP threads used for any given computation will be denoted {\sf\small Nproc}$\times${\sf\small Nthread}.
For a 64 core Xeon Phi, the standard configurations will be  1$\times$64, 2$\times$32, 4$\times$16, 8$\times$8, 16$\times$4, 32$\times$2, and 64$\times$1.   If an application can take advantage of more OpenMP threads than cores, that can easily be set.  For example, 8$\times$32 would have 8 processes each allocating 32 OpenMP threads, nominally consuming 256 cores.  Likewise, for applications where fewer OpenMP threads are optimal, that can also be specified.   For example, 8$\times$2 would have 8 processes each allocating 2 OpenMP threads.  In general the pMatlab manycore syntax makes it very easy to experiment with different combinations of processes and OpenMP threads to find the best performance.  GNU Octave uses the {\sf\small OMP\_NUM\_THREADS} environment variable directly.  For Matlab, additional code is run automatically in a pMatlab launch to align Matlab with {\sf\small OMP\_NUM\_THREADS}

\noindent ~~~~

\noindent ~~~~

----------------------------

{\sf\small Nomp = str2num(getenv({\textquotesingle}OMP\_NUM\_THREADS{\textquotesingle}))

 if (Nomp > 1)

~~~~ maxNumCompThreads(Nomp)
  
 end}

 ----------------------------

There are a variety of patterns that can be used to map OpenMP threads to processor cores.  The {\sf\small KMP\_AFFINITY} environment variable in the Intel compilers can be used to set these patterns \cite{eichenberger2012design}.
For nodes that support hyperthreading, the {\sf\small granularity} modifier  specifies whether to pin OpenMP threads to physical cores ({\sf\small granularity=core}) or logical cores ({\sf\small granularity=fine}).  Using {\sf\small granularity=thread} enables distribution of OpenMP threads in a compact and or scatter fashion \cite{lim2018openmp}.  For this work
$$
{\sf\small KMP\_AFFINITY=granularity=fine}
$$
was used as it prevented  Matlab/Octave from over-allocating OpenMP threads to the same processor core as determined by monitoring the compute node with the Linux {\sf\small htop} command during execution.

\section{Process Pinning}

The Xeon Phi processor employs a memory hierarchy whereby certain tiles, cores, and hyperthreads share different levels of memory.  It can be advantageous to launch processes on the Xeon Phi with an awareness of this memory hierarchy so the underlying OpenMP threads can exploit preferential data locality.  In particular, it is good to avoid having  OpenMP threads execute on cores that are far away from the data they require to operate.  The Linux operating system provides a number of tools for pinning processes to specific logical cores.  This work relies on the {\sf\small taskset --cpu-list} command to launch Matlab/Octave instances that are pinned to specific logical cores.

\begin{figure*}[t]
\centering
\includegraphics[width=2\columnwidth]{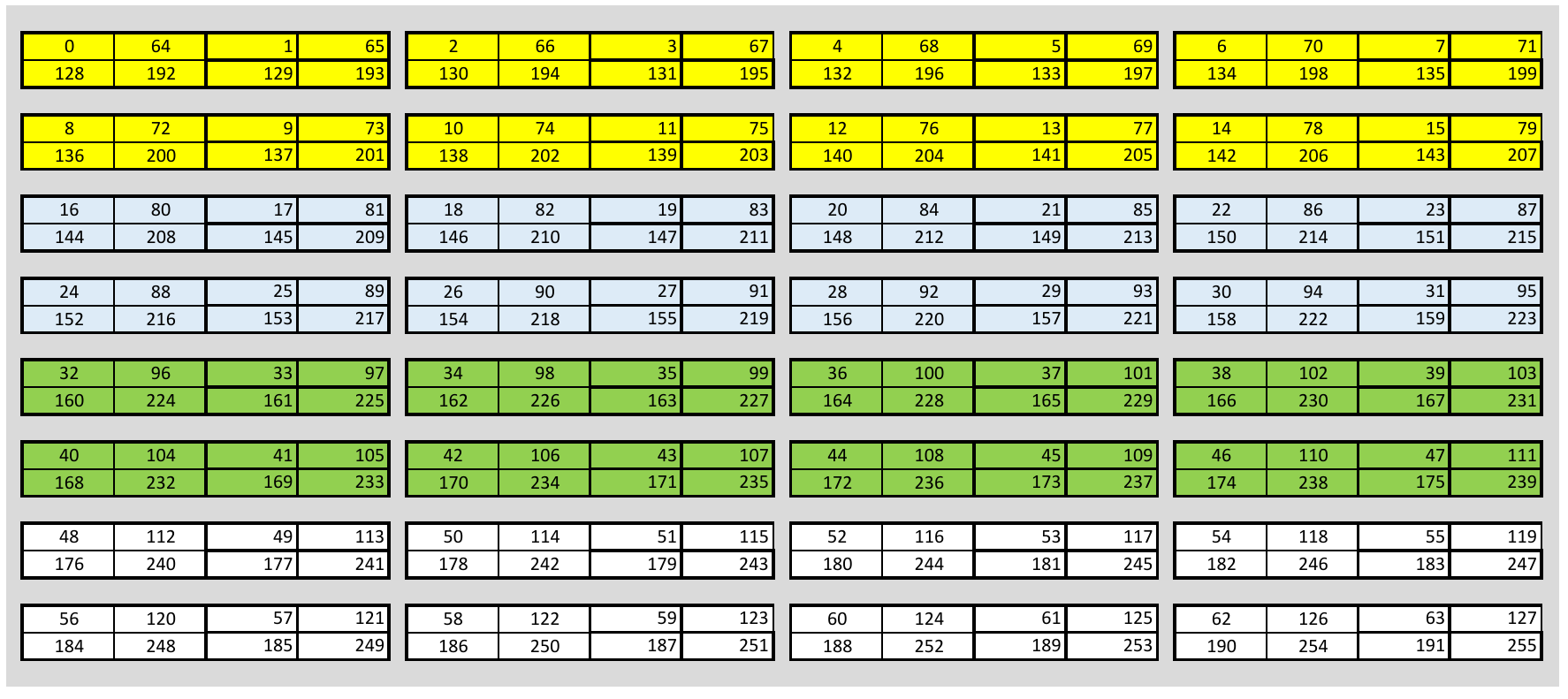}
\caption{Taskset binding of four Matlab/Octave instances (denoted by yellow, blue, green, and white) to the logical core structure of a 32 tile, 64 core, 256 hyperthread Xeon Phi.  This binding maximizes data locality of the underlying OpenMP threads.}
\label{fig:Taskset4processes}
\end{figure*}

The Xeon Phi presents itself to the Linux operating system as 256 cpus (one cpu for each hyperthread).  The cpus $p$, $p$+64, $p$+128, and $p$+192 will be on the same physical processor.  Likewise, if $p$ is even, then cpu $p$+1 will be on the same physical tile.  The mapping of four Matlab/Octave instances to the logical core structure of a 32 tile, 64 core, 256 hyperthread Xeon Phi is illustrated in Figure~\ref{fig:Taskset4processes}. This binding maximizes data locality of the underlying OpenMP threads.

\section{Memory Modes}

Our Xeon Phi processors have two-level memory hierarchy consisting 16 Gigabytes of faster near memory (MCDRAM) and 192 Gigabytes of slower far memory (DRAM) \cite{ramos2017capability,hill2017knl}.  The Xeon Phi has a variety of settings for managing its memory.  These settings are generally set at compute node boot time.

The faster and smaller near memory has three modes: flat, cache, and hybrid. In flat mode both near and far memory form a single address space.  In cache mode the near memory acts as another layer of cache for the far memory.  In hybrid mode, half of the fast memory is flat and half is treated as cache.

The memory can also be divided into different NUMA (non-uniform memory access) domains
\begin{description}
\item[{\sf\small all2all}] cache line addresses are uniformly hashed across the entire memory
\item[{\sf\small hemisphere}] ~~~~~~ cache line addresses are separately hashed into two memory domains
\item[{\sf\small quadrant}] ~~ cache line addresses are separately hashed into four memory domains
\item[{\sf\small snc-2}] sub-NUMA clustering 2 is similar to hemisphere while also exposing each domain for NUMA aware software to exploit
\item[{\sf\small snc-4}] sub-NUMA clustering 4 is similar to quadrant while also exposing each domain for NUMA aware software to exploit
\end{description}

Combined, these combinations of memory modes form 15 distinct configurations
\begin{itemize}
\item {\sf\small all2all-cache}, {\sf\small all2all-flat}, {\sf\small all2all-hybrid}
\item {\sf\small hemisphere-cache}, {\sf\small hemisphere-flat}, {\sf\small hemisphere-hybrid}
\item {\sf\small quadrant-cache}, {\sf\small quadrant-flat}, {\sf\small quadrant-hybrid}
\item {\sf\small snc-2-cache}, {\sf\small snc-2-flat}, {\sf\small snc-2-hybrid}
\item {\sf\small snc-4-cache}, {\sf\small snc-4-flat}, {\sf\small snc-4-hybrid}
\end{itemize}

\section{Performance}

For any particular application, different memory configurations could provide different performance benefits.  The Xeon Phi is designed for vector operations of the  type found in matrix-matrix multiply. Selecting a configuration that is optimal for this operation provides a good foundation for allowing the Xeon Phi to deliver what it was designed to do.  To determine this configuration, 15 Xeon Phi nodes were set in each memory configuration and the Matlab and Octave matrix-matrix multiply performance was measured for various values of {\sf\small Nproc} and {\sf\small Nthread}.

\begin{figure*}[t]
\centering
\includegraphics[width=2\columnwidth]{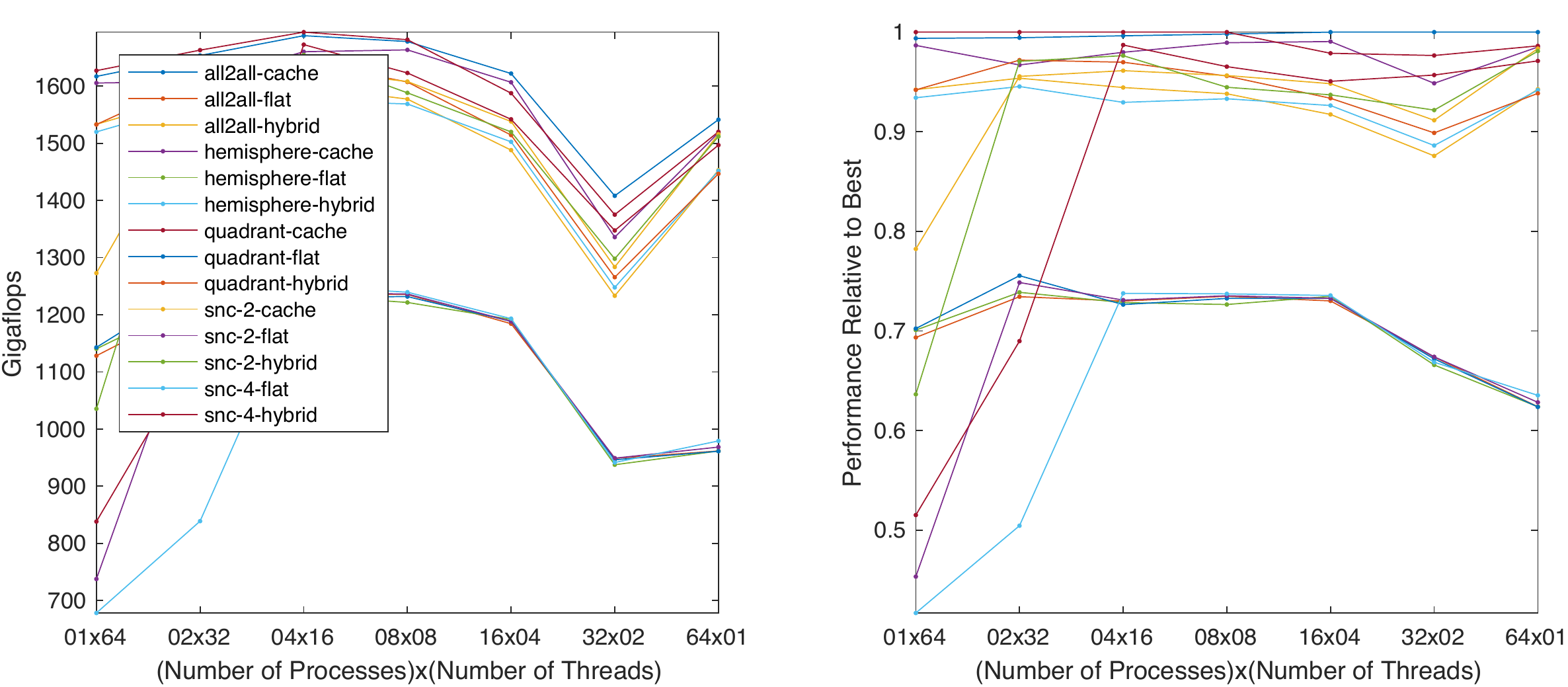}
\includegraphics[width=2\columnwidth]{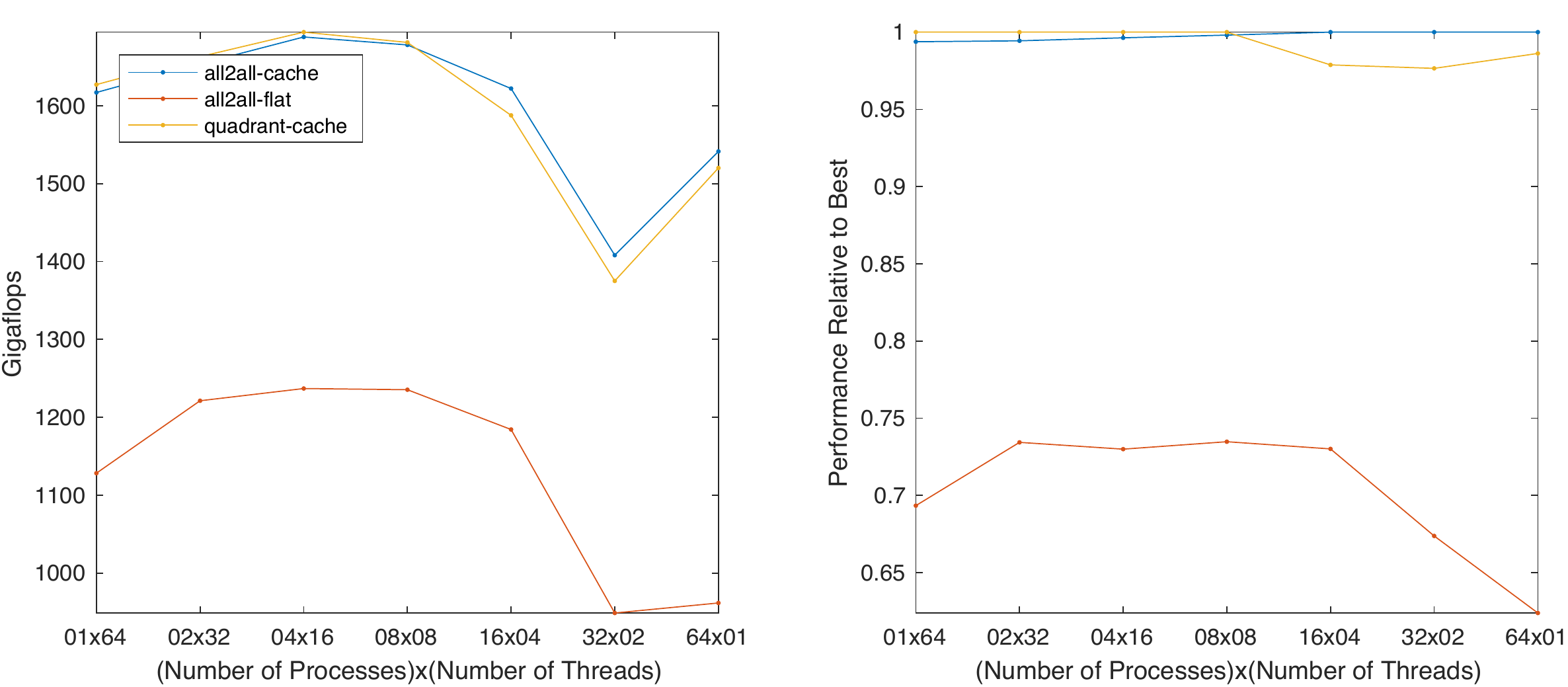}
\caption{[top] Matlab (48000/$\sqrt{{\sf\small Nproc}}$)$\times$(48000/$\sqrt{{\sf\small Nproc}}$) matrix-matrix multiply Gigaflops and relative performance on all memory modes.  [bottom] Gigaflops and relative performance of best performing modes ({\sf\small all2all-cache} and {\sf\small quadrant-cache}) along with the system default ({\sf\small all2all-flat}).}
\label{fig:MatlabPerformance}
\end{figure*}

\begin{figure*}[t]
\centering
\includegraphics[width=2\columnwidth]{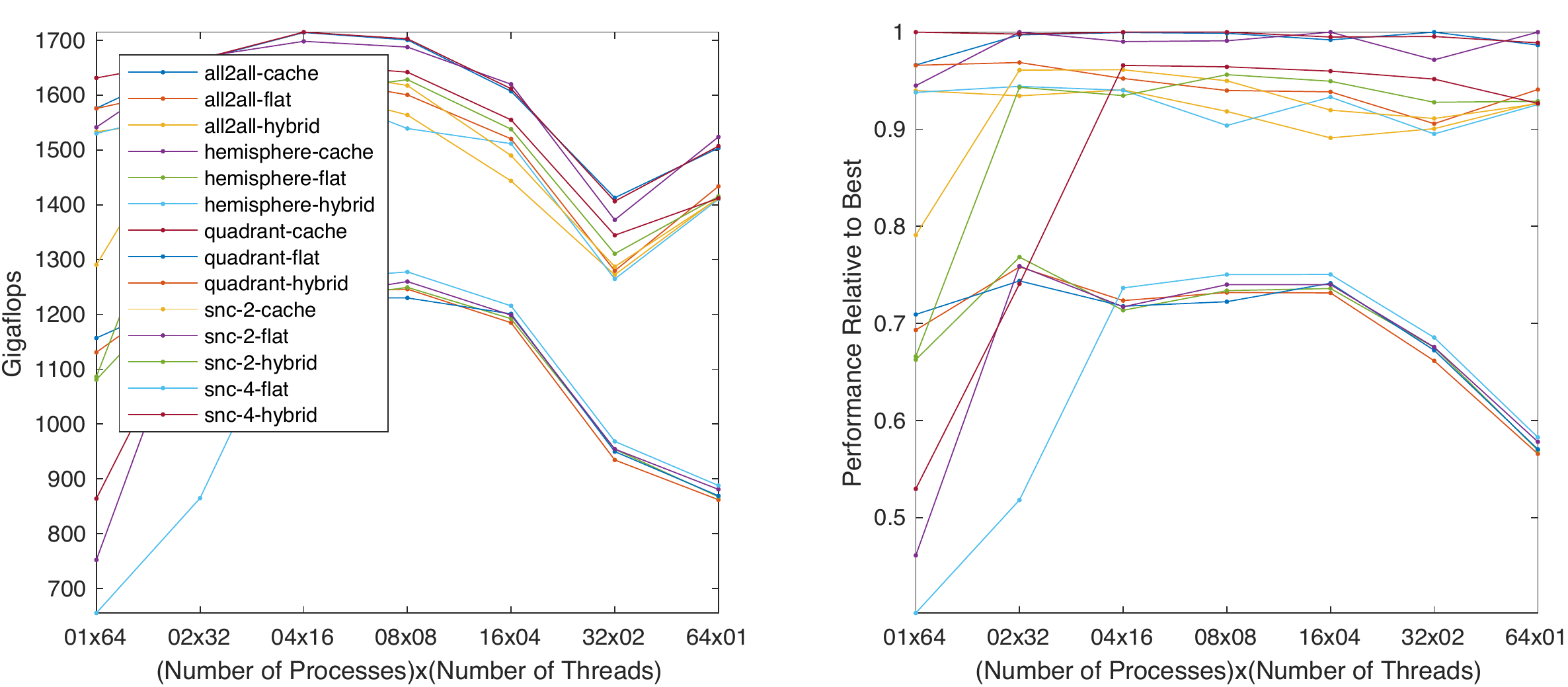}
\includegraphics[width=2\columnwidth]{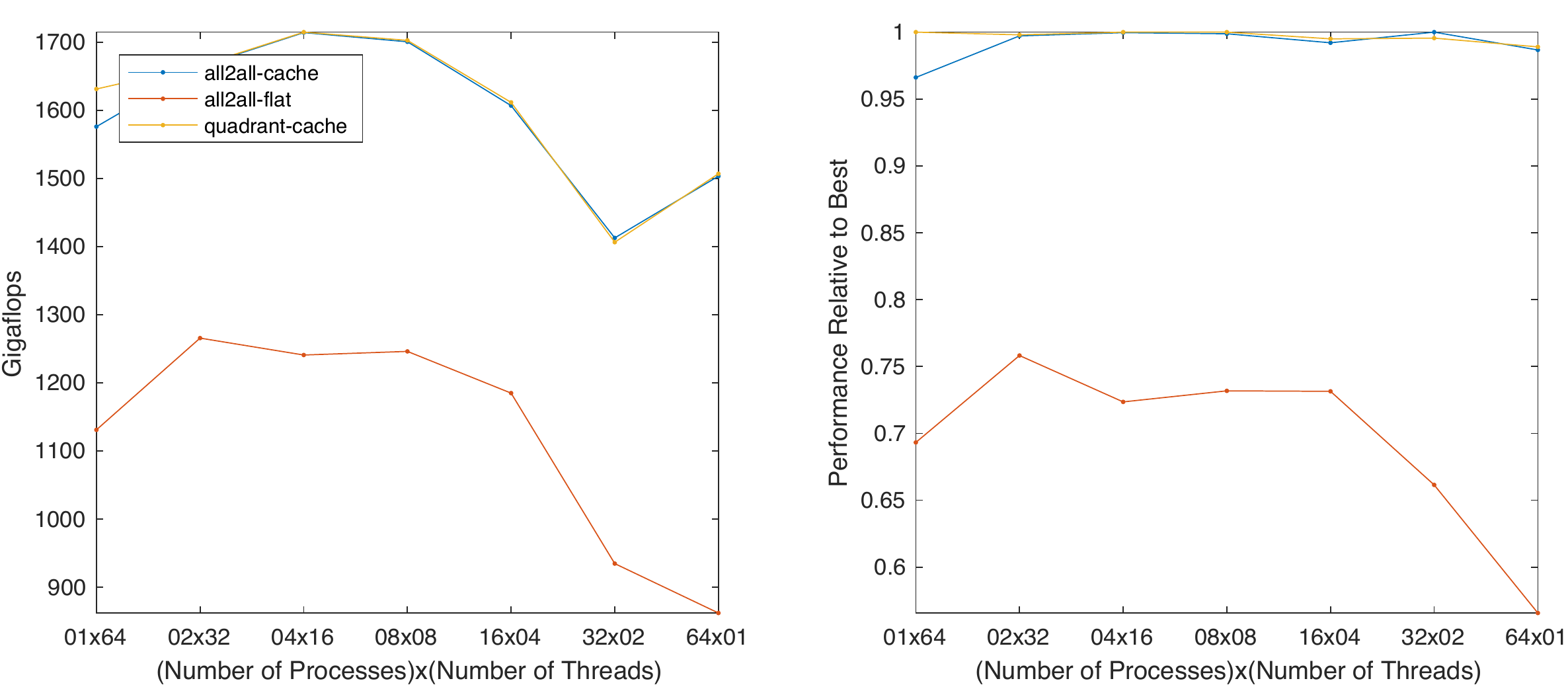}
\caption{[top] GNU Octave (48000/$\sqrt{{\sf\small Nproc}}$)$\times$(48000/$\sqrt{{\sf\small Nproc}}$) matrix-matrix multiply Gigaflops and relative performance on all memory modes.  [bottom] Gigaflops and relative performance of best performing modes ({\sf\small all2all-cache} and {\sf\small quadrant-cache}) along with the system default ({\sf\small all2all-flat}).}
\label{fig:OctavePerformance}
\end{figure*}

  The performance benchmark consisted of each Matlab/Octave instance creating two N$\times$N  matrices $\mathbf{A}$ and $\mathbf{B}$ of random double precision values and multiplying these to form another N$\times$N matrix $\mathbf{C}$.  The total number of bytes required for this operation is 3$\times$8$\times$N$\times$N bytes.  For these experiments the matrix size N was chosen to be 48000/$\sqrt{{\sf\small Nproc}}$ so that the total memory used was the same for all configurations (55 Gigabytes).  The performance results for Matlab version 2018a are shown in Figure~\ref{fig:MatlabPerformance}.  The performance results for GNU Octave version 4.4 are shown in Figure~\ref{fig:OctavePerformance}.  Both Matlab and GNU Octave show similar performance across all memory modes and the performance of the two best modes are ({\sf\small all2all-cache} and {\sf\small quadrant-cache}) are significantly better than the default mode ({\sf\small all2all-flat}).  Based on these data, the LLSC Xeon Phi system selected  {\sf\small all2all-cache} as its default memory mode.
  
   The Xeon Phi 7210 has 128 AVX512 units each capable of performing 16 multiply-accumulate operations per clock cycle.  The AVX512 clock cycle in the Xeon Phi 7210 is 1.1 GHz which means that the practical peak performance is 128$\times$16 (flop) $\times$1.1 GHz = 2252 Gigaflops.  Figures~\ref{fig:MatlabPerformance} and \ref{fig:OctavePerformance} show that a performance of 1500 Gigaflops is consistently achievable, which is 66\% of the practical peak performance of Xeon Phi.

\section{Summary}

The Intel Xeon Phi manycore processor is designed to provide high performance matrix computations of the type often performed in data analysis environments such as Matlab, GNU Octave, Julia, Python, and R. Optimizing the performance of matrix operations within these data analysis environments requires tuning Xeon Phi OpenMP settings, process pinning, and memory modes.  This paper measured matrix-matrix multiplication performance  for Matlab and GNU Octave for different combinations of process counts and OpenMP threads covering all Xeon Phi memory modes.  These measurements indicate that using {\sf\small KMP\_AFFINITY=granlarity=fine}, {\sf\small taskset} pinning, and {\sf\small all2all cache} memory mode allows both Matlab and GNU Octave to achieve 66\% of the practical peak performance of the Xeon Phi.  Using these settings have provided improved performance across a range of applications and has enabled our Xeon Phi system to deliver impactful results on a number  of real-world applications in health sciences \cite{trafton2017mapping}, hurricane relief \cite{foy2018lidar}, astronomy \cite{lindsay2018using}, and cybersecurity \cite{mcgovern2019supercomputers}.

\section*{Acknowledgement}

The authors wish to acknowledge the following individuals for their contributions and support: Bob Bond, Alan Edelman, Charles Leiserson, Dave Martinez, Mimi McClure, Victor Roytburd, and Michael Wright.


%
%




\bibliographystyle{ieeetr}

\bibliography{aarabib}

\begin{thebibliography}{10}

\bibitem{mcmahon1996space}
J.~S. {McMahon} and K.~{Teitelbaum}, ``Space-time adaptive processing on the
  mesh synchronous processor,'' in {\em Proceedings of International Conference
  on Parallel Processing}, pp.~734--740, April 1996.

\bibitem{taylor2002raw}
M.~B. {Taylor}, J.~{Kim}, J.~{Miller}, D.~{Wentzlaff}, F.~{Ghodrat},
  B.~{Greenwald}, H.~{Hoffman}, P.~{Johnson}, {Jae-Wook Lee}, W.~{Lee},
  A.~{Ma}, A.~{Saraf}, M.~{Seneski}, N.~{Shnidman}, V.~{Strumpen}, M.~{Frank},
  S.~{Amarasinghe}, and A.~{Agarwal}, ``The raw microprocessor: a computational
  fabric for software circuits and general-purpose programs,'' {\em IEEE
  Micro}, vol.~22, pp.~25--35, March 2002.

\bibitem{mattson2008programming}
T.~G. Mattson, R.~Van~der Wijngaart, and M.~Frumkin, ``Programming the intel
  80-core network-on-a-chip terascale processor,'' in {\em Proceedings of the
  2008 ACM/IEEE Conference on Supercomputing}, SC '08, (Piscataway, NJ, USA),
  pp.~38:1--38:11, IEEE Press, 2008.

\bibitem{ramey2011tile}
C.~{Ramey}, ``Tile-gx100 manycore processor: Acceleration interfaces and
  architecture,'' in {\em 2011 IEEE Hot Chips 23 Symposium (HCS)}, pp.~1--21,
  Aug 2011.

\bibitem{sodani2015knights}
A.~{Sodani}, ``Knights landing (knl): 2nd generation intel® xeon phi
  processor,'' in {\em 2015 IEEE Hot Chips 27 Symposium (HCS)}, pp.~1--24, Aug
  2015.

\bibitem{sodani2016knights}
A.~{Sodani}, R.~{Gramunt}, J.~{Corbal}, H.~{Kim}, K.~{Vinod}, S.~{Chinthamani},
  S.~{Hutsell}, R.~{Agarwal}, and Y.~{Liu}, ``Knights landing:
  Second-generation intel xeon phi product,'' {\em IEEE Micro}, vol.~36,
  pp.~34--46, Mar 2016.

\bibitem{kepnerjananthan}
J.~Kepner and H.~Jananthan, {\em Mathematics of Big Data}.
\newblock MIT Press, 2018.

\bibitem{dongarra2015hpc}
J.~Dongarra, M.~Gates, A.~Haidar, Y.~Jia, K.~Kabir, P.~Luszczek, and S.~Tomov,
  ``Hpc programming on intel many-integrated-core hardware with magma port to
  xeon phi,'' {\em Sci. Program.}, vol.~2015, pp.~9:9--9:9, Jan. 2015.

\bibitem{doerfler2016roofline}
D.~Doerfler, J.~Deslippe, S.~Williams, L.~Oliker, B.~Cook, T.~Kurth, M.~Lobet,
  T.~Malas, J.-L. Vay, and H.~Vincenti, ``Applying the roofline performance
  model to the intel xeon phi knights landing processor,'' in {\em High
  Performance Computing} (M.~Taufer, B.~Mohr, and J.~M. Kunkel, eds.), (Cham),
  pp.~339--353, Springer International Publishing, 2016.

\bibitem{haidar2016lu}
A.~{Haidar}, S.~{Tomov}, K.~{Arturov}, M.~{Guney}, S.~{Story}, and
  J.~{Dongarra}, ``Lu, qr, and cholesky factorizations: Programming model,
  performance analysis and optimization techniques for the intel knights
  landing xeon phi,'' in {\em 2016 IEEE High Performance Extreme Computing
  Conference (HPEC)}, pp.~1--7, Sep. 2016.

\bibitem{jeffers2016intel}
J.~Jeffers, J.~Reinders, and A.~Sodani, {\em Intel Xeon Phi Processor High
  Performance Programming: Knights Landing Edition}.
\newblock Morgan Kaufmann, 2016.

\bibitem{chunduri2017analytical}
S.~Chunduri, P.~Balaprakash, V.~Morozov, V.~Vishwanath, and K.~Kumaran,
  ``Analytical performance modeling and validation of intel's xeon phi
  architecture,'' in {\em Proceedings of the Computing Frontiers Conference},
  CF'17, (New York, NY, USA), pp.~247--250, ACM, 2017.

\bibitem{nagasaka2018high}
Y.~Nagasaka, S.~Matsuoka, A.~Azad, and A.~Bulu{\c{c}}, ``High-performance
  sparse matrix-matrix products on intel knl and multicore architectures,''
  {\em arXiv preprint arXiv:1804.01698}, 2018.

\bibitem{lim2018implementation}
R.~Lim, Y.~Lee, R.~Kim, and J.~Choi, ``An implementation of matrix--matrix
  multiplication on the intel knl processor with avx-512,'' {\em Cluster
  Computing}, vol.~21, pp.~1785--1795, Dec 2018.

\bibitem{kepner2018design}
J.~{Kepner}, S.~{Samsi}, W.~{Arcand}, D.~{Bestor}, B.~{Bergeron}, T.~{Davis},
  V.~{Gadepally}, M.~{Houle}, M.~{Hubbell}, H.~{Jananthan}, M.~{Jones},
  A.~{Klein}, P.~{Michaleas}, R.~{Pearce}, L.~{Milechin}, J.~{Mullen},
  A.~{Prout}, A.~{Rosa}, G.~{Sanders}, C.~{Yee}, and A.~{Reuther}, ``Design,
  generation, and validation of extreme scale power-law graphs,'' in {\em 2018
  IEEE International Parallel and Distributed Processing Symposium Workshops
  (IPDPSW)}, pp.~279--286, May 2018.

\bibitem{gadepally2018hyperscaling}
V.~{Gadepally}, J.~{Kepner}, L.~{Milechin}, W.~{Arcand}, D.~{Bestor},
  B.~{Bergeron}, C.~{Byun}, M.~{Hubbell}, M.~{Houle}, M.~{Jones},
  P.~{Michaleas}, J.~{Mullen}, A.~{Prout}, A.~{Rosa}, C.~{Yee}, S.~{Samsi}, and
  A.~{Reuther}, ``Hyperscaling internet graph analysis with d4m on the mit
  supercloud,'' in {\em 2018 IEEE High Performance extreme Computing Conference
  (HPEC)}, pp.~1--6, Sep. 2018.

\bibitem{jones2018interactive}
M.~{Jones}, J.~{Kepner}, B.~{Orchard}, A.~{Reuther}, W.~{Arcand}, D.~{Bestor},
  B.~{Bergeron}, C.~{Byun}, V.~{Gadepally}, M.~{Houle}, M.~{Hubbell},
  A.~{Klein}, L.~{Milechin}, J.~{Mullen}, A.~{Prout}, A.~{Rosa}, S.~{Samsi},
  C.~{Yee}, and P.~{Michaleas}, ``Interactive launch of 16,000 microsoft
  windows instances on a supercomputer,'' in {\em 2018 IEEE High Performance
  extreme Computing Conference (HPEC)}, pp.~1--6, Sep. 2018.

\bibitem{reuther2018interactive}
A.~{Reuther}, J.~{Kepner}, C.~{Byun}, S.~{Samsi}, W.~{Arcand}, D.~{Bestor},
  B.~{Bergeron}, V.~{Gadepally}, M.~{Houle}, M.~{Hubbell}, M.~{Jones},
  A.~{Klein}, L.~{Milechin}, J.~{Mullen}, A.~{Prout}, A.~{Rosa}, C.~{Yee}, and
  P.~{Michaleas}, ``Interactive supercomputing on 40,000 cores for machine
  learning and data analysis,'' in {\em 2018 IEEE High Performance extreme
  Computing Conference (HPEC)}, pp.~1--6, Sep. 2018.

\bibitem{kepner2018tabularosa}
J.~{Kepner}, R.~{Brightwell}, A.~{Edelman}, V.~{Gadepally}, H.~{Jananthan},
  M.~{Jones}, S.~{Madden}, P.~{Michaleas}, H.~{Okhravi}, K.~{Pedretti},
  A.~{Reuther}, T.~{Sterling}, and M.~{Stonebraker}, ``Tabularosa: Tabular
  operating system architecture for massively parallel heterogeneous compute
  engines,'' in {\em 2018 IEEE High Performance extreme Computing Conference
  (HPEC)}, pp.~1--8, Sep. 2018.

\bibitem{trafton2017mapping}
A.~Trafton, ``Mapping the brain, cell by cell,'' {\em MIT News}, 2018.

\bibitem{foy2018lidar}
K.~Foy, ``Lidar accelerates hurricane recovery in the carolinas,'' {\em MIT
  News}, 2018.

\bibitem{lindsay2018using}
R.~Lindsay, ``Using lidar to assess destruction in puerto rico,'' {\em MIT
  News}, 2018.

\bibitem{mcgovern2019supercomputers}
A.~McGovern, ``Supercomputers can spot cyber threats,'' {\em MIT News}, 2019.

\bibitem{kepner2009parallel}
J.~Kepner, {\em Parallel MATLAB for multicore and multinode computers},
  vol.~21.
\newblock SIAM, 2009.

\bibitem{eichenberger2012design}
A.~E. Eichenberger, C.~Terboven, M.~Wong, and D.~an~Mey, ``The design of openmp
  thread affinity,'' in {\em OpenMP in a Heterogeneous World} (B.~M. Chapman,
  F.~Massaioli, M.~S. M{\"u}ller, and M.~Rorro, eds.), (Berlin, Heidelberg),
  pp.~15--28, Springer Berlin Heidelberg, 2012.

\bibitem{lim2018openmp}
R.~Lim, Y.~Lee, R.~Kim, and J.~Choi, ``Openmp-based parallel implementation of
  matrix-matrix multiplication on the intel knights landing,'' in {\em
  Proceedings of Workshops of HPC Asia}, HPC Asia '18, (New York, NY, USA),
  pp.~63--66, ACM, 2018.

\bibitem{ramos2017capability}
S.~{Ramos} and T.~{Hoefler}, ``Capability models for manycore memory systems: A
  case-study with xeon phi knl,'' in {\em 2017 IEEE International Parallel and
  Distributed Processing Symposium (IPDPS)}, pp.~297--306, May 2017.

\bibitem{hill2017knl}
P.~Hill, C.~Snyder, and J.~Sygulla, ``Knl system software,'' {\em Cray User
  Group CUG, May}, 2017.

\end{thebibliography}
%

\end{document}